\documentclass[12pt,preprint]{aastex}

\shorttitle{Radiation Pressure-supported Disks}
\shortauthors{Gu}

\newcommand{\bma}[1]{\mbox{\boldmath $#1$}}

\begin{document}

\title{Radiation Pressure-supported Accretion Disks:
Vertical Structure, Energy Advection, and Convective Stability}

\author{Wei-Min Gu\altaffilmark{1,2}}

\altaffiltext{1}{Department of Physics and Institute of Theoretical Physics
and Astrophysics, Xiamen University, Xiamen, Fujian 361005, China}

\altaffiltext{2}{Harvard-Smithsonian Center for Astrophysics,
60 Garden Street, Cambridge, MA 02138, USA}

\email{guwm@xmu.edu.cn}

\begin{abstract}
By taking into account the local energy balance per unit volume between
the viscous heating and the advective cooling plus the radiative cooling,
we investigate the vertical structure of radiation pressure-supported
accretion disks in spherical coordinates.
Our solutions show that the photosphere of the disk is close to
the polar axis and therefore the disk seems to be extremely thick.
However, the profile of density implies that most of the accreted
matter exists in a moderate range around the equatorial plane.
We show that the well-known polytropic relation between the pressure and
the density is unsuitable for describing the vertical structure
of radiation pressure-supported disks.
More importantly, we find that the energy advection is significant
even for slightly sub-Eddington accretion disks.
We argue that the non-negligible advection may help to understand
why the standard thin disk model is likely to be inaccurate above
$\sim 0.3$ Eddington luminosity, which
was found by some works on the black hole spin measurement.
Furthermore, the solutions satisfy the Solberg-H\o{}iland
conditions, which indicates the disk to be convectively stable.
In addition, we discuss the possible link between our disk model
and ultraluminous X-ray sources.
\end{abstract}


\section{Introduction}
The standard thin accretion disk model \citep{SS73}
has been widely applied to X-ray binaries and active galactic
nuclei. Due to the basic assumption of the energy balance between
the viscous heating and the radiative cooling, such a model was known
to be invalid for super-Eddington accretion case, where the advective
cooling is probably significant. Instead, the slim disk model \citep{Abram88}
was introduced to describe super-Eddington accretion
disks. However, there exists some conflict between the theory
and the observation. The theory predicts that the advection is negligible
for $L \la L_{\rm Edd}$ \citep[e.g.,][]{Watarai00,Sadow11a},
where $L_{\rm Edd}$ is the Eddington luminosity,
which indicates that the standard
disk model should be valid up to $L_{\rm Edd}$.
On the contrary, some works on the black hole spin measurement
showed that the standard disk model is likely to be inaccurate
for $L \ga 0.3 L_{\rm Edd}$
\citep[e.g.,][]{McClin06}. Moreover, even the recent general model for
optically thick disks \citep[e.g.,][]{Sadow11a,Sadow11b},
which unifies the standard thin disk and the slim disk,
could not help to obtain a self-consistent spin parameter
for $L \ga 0.3 L_{\rm Edd}$ \citep[e.g.,][]{Straub11}.
In our opinion, the above conflict may be resolved if the vertical
structure is well incorporated.

Most previous works on accretion disks focused on the radial structure
in cylindrical coordinates ($R$, $\phi$, $z$).
For the vertical structure, however,
a simple well-known relationship ``$H = c_{\rm s}/\Omega_{\rm K}$" or
``$H \Omega_{\rm K}/c_{\rm s} = {\rm constant}$" was widely
adopted, where $H$ is the half-height of the disk, $c_{\rm s}$ is
the sound speed, and $\Omega_{\rm K}$ is the Keplerian angular velocity.
Such a relationship comes from the vertical hydrostatic equilibrium
with two additional assumptions.
One is the approximation of gravitational potential:
$\psi(R,z) \simeq \psi(R,0) + \Omega_{\rm K}^2 z^2/2$, and the other
is a one-zone approximation or a polytropic relation
$p_{\rm tot} = {\mathcal K} \rho^{1+1/N}$ in the vertical direction
\citep[e.g.,][]{Hoshi77},
where $p_{\rm tot}$ is the total (gas plus radiation) pressure
and $\rho$ is the density.
Obviously, the above assumptions work well for
geometrically thin disks, but may be inaccurate for the mass accretion
rate $\dot M$ approaching the Eddington one $\dot M_{\rm Edd}$,
for which the disk is probably not thin.
Consequently, the relationship
``$H \Omega_{\rm K}/c_{\rm s} = {\rm constant}$" may be invalid for
$\dot M \ga \dot M_{\rm Edd}$.

Without the potential approximation, our two previous works investigated
the geometrical thickness of accretion disks and the validity of
the relationship ``$H \Omega_{\rm K}/c_{\rm s} = {\rm constant}$".
\citet{Gu07} adopted the explicit gravitational potential
in cylindrical coordinates and found that the above relationship is
inaccurate for $\dot M \ga \dot M_{\rm Edd}$, and therefore the disk
can be geometrically thick. \citet{Gu09} took spherical
coordinates to avoid the approximation of gravitational potential, and
found that an advection-dominated accretion disk is likely to be
quite thick. In these two works, however, the polytropic relation
is still adopted in the vertical direction,
which takes the place of the energy balance per unit volume between
the viscous heating and the advective cooling plus the radiative cooling.
The validity of such a polytropic relation, however, remains questionable,
in particular for large $\dot M$ due to dominant radiation pressure.

The purpose of this paper is to revisit the vertical structure of
radiation pressure-supported disks by taking into account
the local energy balance and to study the variation of
energy advection with mass accretion rates.
The paper is organized as follows.
Equations and boundary conditions are derived in Section~2.
A global view of the solutions in the $\dot m$-$r$ diagram
is presented in Section~3. For a typical radius $r = 10 r_{\rm g}$,
the vertical structure and the energy advection are investigated in Section~4.
The two-dimensional solutions and the convective stability are studied
in Section~5. Summary and discussion are made in Section~6.

\section{Equations and boundary conditions}

\subsection{Equations}

We consider a steady state axisymmetric accretion disk in
spherical coordinates ($r$, $\theta$, $\phi$) and use the Newtonian
potential, $\psi = -GM/r$, where $M$ is the black hole mass.
Following \citet{NY95}, we assume $v_{\theta} = 0$ for
simplicity, which means a hydrostatic equilibrium in the $\theta$ direction.
Simulations, \citep[e.g.,][Figure~3]{Ohsuga05}, however, revealed that
$v_{\theta}$ will be significant for extremely high accretion rates
such as $\dot M = 1000 L_{\rm Edd}/c^2$.
As shown in the following sections, our solutions mainly correspond
to $\dot M$ around $\dot M_{\rm Edd}$. For such accretion rates,
the validity of $v_{\theta} = 0$ remains a question.

The basic equations of continuity and momentum take the forms
\citep[e.g.,][]{Kato08}:
\begin{equation}
\frac{1}{r^2} \frac{\partial}{\partial r} (r^2 \rho v_r) = 0 \ ,
\end{equation}
\begin{equation}
v_r \frac {\partial v_r}{\partial r} - \frac{v_{\phi}^2}{r}
= - \frac{GM}{r^2} - \frac{1}{\rho} \frac{\partial p}{\partial r}
+ \frac{\kappa_{\rm es}}{c} F_r \ ,
\end{equation}
\begin{equation}
- \frac{v_{\phi}^2}{r} \cot \theta = - \frac{1}{\rho r}
\frac{\partial p}{\partial \theta} + \frac{\kappa_{\rm es}}{c} F_{\theta} \ ,
\end{equation}
\begin{equation}
\frac{v_r}{r} \frac{\partial}{\partial r} (r v_{\phi})
= \frac{1}{\rho r^3} \frac{\partial}{\partial r} (r^3 \tau_{r\phi}) \ ,
\end{equation}
where $v_r$ and $v_{\phi}$ are respectively the radial and azimuthal
velocity, $F_r$ and $F_{\theta}$ are respectively the radial and
vertical radiation flux, $p$ is the gas pressure,
$\kappa_{\rm es}$ is the opacity of electron scattering,
and $\tau_{r\phi}$ is the $r\phi$ component of the viscous stress tensor,
$\tau_{r\phi} = \nu \rho r \partial (v_{\phi}/r) /\partial r$.
Following the spirit of $\alpha$ stress prescription,
we assume the kinematic viscosity coefficient
$\nu = \alpha c_{\rm s}^2 r / v_{\rm K}$, where $c_{\rm s}$ is the sound
speed defined below (Equation~(7)), and $v_{\rm K} = (GM/r)^{1/2}$
is the Keplerian velocity.

We would stress that, even though the $\alpha$ stress prescription
has been widely adopted for theoretical studies, simulations of
magnetorotational turbulence have shown that the stress does not
well scale locally with the pressure. For instance, the simulations
on thin disks by a shearing box showed that, the time and box-averaged
results are likely to support that the stress is proportional to the
thermal (gas plus radiation) pressure
\citep[e.g.,][Figure~3]{Hirose09a}.
However, Figure~16 of \citet{Hirose09b} shows that
the maximal thermal pressure is located on
the equatorial plane, whereas Figure~11 shows that
the maximal stress is obviously not at the same place.
These two figures reveal that the stress is not
proportional to the pressure locally.
In the present study, for simplicity, we will keep
the local $\alpha$ stress prescription for numerical calculation,
which is a weak point of this work.

The energy equation including gas and radiation is written as
\citep[e.g.,][]{Ohsuga05}
\begin{equation}
{\bma \nabla} \cdot [(e+E) {\bma v}] = -p {\bma \nabla} \cdot {\bma v}
- {\bma \nabla} {\bma v} {\bma :} {\bma P} - {\bma \nabla} \cdot {\bma F}
+ \Phi_{\rm vis} \ ,
\end{equation}
where $e$ and $E$ are the internal energy density of the gas
and the radiation, respectively. ${\bma P} = \bma{f} E$ is the radiation
pressure tensor, $\Phi_{\rm vis}$ is the viscous dissipative function,
and the radiation flux ${\bma F}$ is expressed as
\begin{equation}
{\bma F} = - \frac{\lambda c}{\rho \kappa_{\rm es}} {\bma \nabla} E \ .
\end{equation}
In this work, we focus on the region inside the photosphere, so we can
take the well-known Eddington approximation, i.e.,
$\lambda = 1/3$ and the Eddington tensor ${\bma f} = {\bma I}/3$.

Since we only study the radiation pressure-supported disks, the gas
pressure $p$ and the gas internal energy density $e$ will be
dropped in our calculation.
In order to avoid directly solving the partial differential equations, some
assumptions on the radial derivatives ($\partial/\partial r$) are required.
Following the spirit of self-similar assumptions
\citep[e.g.,][]{BM82,NY95},
we adopt the following radial derivatives for $c_{\rm s}$ and $E$:
\[
\frac{\partial \ln c_{\rm s}}{\partial \ln r} = - \frac{1}{2} \ ;
\qquad
\frac{\partial \ln E}{\partial \ln r} = - \frac{5}{2} \ ,
\]
where the sound speed $c_{\rm s}$ is defined as
\begin{equation}
c_{\rm s}^2 \equiv \frac{E}{3 \rho} \ .
\end{equation}
Based on the above two radial derivatives,
the following four derivatives can be inferred from Equations~(1)-(7): 
\[
\frac{\partial \ln |v_r|}{\partial \ln r} = - \frac{1}{2} \ ; \
\frac{\partial \ln v_{\phi}}{\partial \ln r} = - \frac{1}{2} \ ; \
\frac{\partial \ln \rho}{\partial \ln r} = - \frac{3}{2} \ ; \
\frac{\partial \ln F_r}{\partial \ln r} = -2 \ .
\]

With all the above derivatives, we can remove the ``$\partial/\partial r$"
terms in Equations (2) and (4)-(6), and the following equations are
then obtained from Equations (2)-(6):
\begin{equation}
\frac{1}{2} v_r^2 + \frac{5}{2} c_{\rm s}^2 + v_{\phi}^2 - v_{\rm K}^2 = 0 \ ,
\end{equation}
\begin{equation}
v_{\phi}^2 \cot\theta = - \frac{r \kappa_{\rm es}}{c} F_{\theta} \ ,
\end{equation}
\begin{equation}
v_r = - \frac{3}{2} \frac{\alpha c_{\rm s}^2}{v_{\rm K}} \ ,
\end{equation}
\begin{equation}
- \frac{1}{2} v_r (3 \rho v_{\phi}^2 - E)
= \frac{1}{\sin\theta} \frac{d}{d \theta} (\sin\theta F_{\theta}) \ ,
\end{equation}
\begin{equation}
\frac{d E}{d \theta} = - \frac{3 r \rho \kappa_{\rm es}}{c} F_{\theta} \ ,
\end{equation}
\begin{equation}
F_r = \frac{5}{6} \frac{c E}{r \rho \kappa_{\rm es}} \ .
\end{equation}

The seven equations, Equations~(7)-(13), enable us to solve
for the seven variables: $v_r$, $v_{\phi}$, $c_{\rm s}$, $\rho$,
$E$, $F_r$, and $F_{\theta}$.
There are two differential equations in this system. In addition,
the position of the surface is unknown. Thus, totally three boundary
conditions are required to determine a unique solution. 

\subsection{Boundary conditions}

An obvious boundary condition on the equatorial plane is $F_{\theta} = 0$.
However, this condition is not applicable for numerical calculation
since it is automatically matched as indicated by Equation~(9).
Combining Equations (9) and (11) we can derive the following equation:
\[
\cot\theta \frac{d}{d \theta} (v_{\phi}^2)
= v_{\phi}^2 + \frac{r v_r \kappa_{\rm es}}{2 c} (3 \rho v_{\phi}^2 - E) \ .
\]
An alternative boundary condition on the equatorial plane
is then obtained from the above equation (the left-hand
side is zero thus the right-hand side should also be zero):
\begin{equation}
v_{\phi}^2 + \frac{r v_r \kappa_{\rm es}}{2 c} (3 \rho v_{\phi}^2 - E)
= 0 \ \ \ (\theta = \frac{\pi}{2})  \ .
\end{equation}
The second boundary condition is a definition of the surface.
We define the photosphere as the position above which
the optical depth is around unity. The condition can be written as
\begin{equation}
\tau_{\rm es} \equiv
r \kappa_{\rm es} \rho^2 \left( \frac{d \rho}{d \theta} \right)^{-1} = 1
\ \ \ (\theta = \theta_0) \ ,
\end{equation}
where $\theta_0$ ($0 < \theta_0 < \pi/2$) is the polar angle of
the photosphere.
The third condition is related to the mass accretion rate:
\begin{equation}
\dot M = - 2 \pi r^2 \int_{\theta_0}^{\pi-\theta_0} \rho v_r \sin \theta
\ d\theta \ .
\end{equation}

\section{Solutions in $\dot m$-$r$ diagram}

In our calculation we set $M = 10 M_{\sun}$, $\kappa_{\rm es} =
0.34$ cm$^2$g$^{-1}$, and $\alpha = 0.02$, where the value of $\alpha$
is taken from recent simulations \citep{Hirose09a}.
The Eddington accretion rate is expressed as
$\dot M_{\rm Edd} = 4\pi GM / \eta c\kappa_{\rm es}$, where $\eta$ is
a radiative efficiency of the flow. We choose
$\eta = 1/16$ since it is comparable to the Schwarzschild black hole
efficiency of 0.057. The dimensionless
accretion rate is defined as $\dot m \equiv \dot M / \dot M_{\rm Edd}$.

With the equations and boundary conditions in Section 2, we can numerically
derive the $\theta$-direction distribution of physical quantities
for a given $\dot m$ at a certain radius $r$. The radiation
pressure-supported disk solutions in the $\dot m$-$r$ diagram are
shown in Figure~1, where $r_{\rm g} \equiv 2GM/c^2$ is the Schwarzschild radius.
The parameter space is divided into three regions
by two parallel solid lines, roughly with $\dot m \propto r$.
The region above the upper solid line is denoted by ``Outflow",
where we cannot find solutions.
No solution exists probably due to the assumption of $v_{\theta}=0$ in advance.
In our view, a real flow located in this region may have
$v_{\theta} \ne 0$ and the inflow accretion rate may decrease inward.
The physical understanding could be that, for high accretion rates
and particularly for the inner radii, the viscous dissipation
may be sufficiently large such that the radiation pressure is too
strong to be balanced by the gravitational force. Thus, outflows may
be driven by the radiation pressure and the inflow $\dot m$ drops inward.
On the other hand, simulations of supercritical accretion flows
\citep[e.g.,][Figure~6]{Ohsuga05} showed that the inflow accretion rate
roughly follows the $\dot m \propto r$ relationship for
$\dot m = 1000 L_{\rm Edd}/c^2$ at $r_{\rm out} = 500 r_{\rm g}$
(corresponding to $\dot m = 62.5$
due to the definition of $\dot M_{\rm Edd}$ with $\eta = 1/16$).
The slope of the upper solid line in Figure~1, which may be regarded as
maximal accretion rates due to our calculation,
agrees well with the slope in the above simulations.

The region under the lower solid line is denoted by ``Gas pressure",
where no solution is found either.
In our understanding, it is probably because the gas pressure
cannot be ignored in this region, which may be in conflict with
the radiation pressure-supported assumption. We would point out
that, the lower solid line in this diagram is higher than the well-known
line which separates the inner and middle regions of standard thin disks
\citep{SS73}. The reason is that, the gas and radiation
pressure are comparable for the latter, whereas the
radiation pressure-supported disk may require the accretion rate
to be higher such that the radiation pressure sufficiently dominate over
the gas pressure, and therefore the effect of gas pressure on
the vertical structure can be completely ignored.

The region between the two solid line, denoted by ``Radiation pressure",
which means that the radiation pressure is completely dominated,
corresponds to the solutions of our main interest in this work.
In Section~4, we will focus on the vertical structure and the energy
advection at a typical radius, $r=10 r_{\rm g}$, as indicated
by the vertical dashed line in Figure~1. In Section~5, we will study the
two-dimensional solutions for a typical accretion rate $\dot m = 0.6$
in the range $6 r_{\rm g} \leqslant r \leqslant 12 r_{\rm g}$ and
$0 < \theta \leqslant \pi/2$, as indicated by the horizontal dot-dashed line.
In addition, we would point out that for inner radii such as
$3\sim 5 r_{\rm g}$, the two solid lines in Figure~1
may deviate from a real black hole accretion system
due to the Newtonian potential used in this work.

\section{Solutions at a typical radius $r = 10 r_{\rm g}$}

\subsection{Vertical structure}

In this section, we will focus on the solutions at a typical radius
$r = 10r_{\rm g}$.
Figure~2 shows the vertical structure of the disk with $\dot m = 0.6$.
In Figure~2(a), the dot-dashed, dotted, solid, and dashed lines
show the vertical distribution of the dimensionless density ($\rho/\rho_0$),
radial velocity ($v_r/v_{\rm K}$), azimuthal velocity
($v_{\phi}/v_{\rm K}$), and sound speed ($c_{\rm s}/v_{\rm K}$),
respectively, where $\rho_0$ is the density on the equatorial plane.
It is seen that $\rho$ significantly decreases, whereas $c_{\rm s}$ and
$|v_r|$ increases, from the equatorial plane to the surface.
In Figure~2(b), the solid line shows the variation of $\tau_{\rm es}$
(defined in Equation~(15)), where the photosphere ($\tau_{\rm es} =1$) is
located at $\theta_0 \approx 4^{\circ}$, quite close to
the polar axis. The disk seems to be extremely thick according to
the position of the photosphere. However, the profile of $\rho$ implies
that most of the accreted matter exists in a moderate range around the
equatorial plane, such as $\pi/4 < \theta < 3\pi/4$, which is more
clear in Figures~4 and 5 (discussed below). 
The dashed line shows the variation of $|F_{\theta}|/cE$.
There exists $|F_{\theta}|/cE \la 1/3$ for the whole solution,
which indicates that the Eddington approximation is valid
and the solution is therefore self-consistent.

With a more general viscosity, \citet{BM82} studied a geometrically thick,
radiation pressure-supported model for supercritical accretion disks.
They showed that there exists a narrow empty funnel along the rotation axis
with a half-opening angle $\la 4^{\circ}.6$. As seen in our Figure~2(a),
the density drops sharply close to the photosphere, thus a nearly
empty funnel also seems to exist in our model. The difference
is that, the disk surface in \citet{BM82} is the position where
some physical quantities such as $v_r$ diverges,
whereas our surface is defined as the position where Equation~(15)
is matched, and no divergence appears in our solutions.

For a real disk with $\dot m = 0.6$,
the photosphere may exist between $\theta = 45^{\circ}$ and the present
result ($\approx 4^{\circ}$). Our argument is as follows.
There are two possible reasons that may cause the present
photosphere quite close to the polar axis. One is that we have ignored the
radiation force from one side (e.g., $\theta = \theta_0$ and $0 < \phi < \pi$)
to the other (e.g., $\theta = \theta_0$ and $\pi < \phi < 2\pi$).
The other reason is that we consider only the $r\phi$ component of
the stress tensor, which may cause inaccurate results for
small $\theta$, such as strong shearing of the angular velocity
$\Omega$ in the vertical direction, where $\Omega = v_{\phi}/(r\sin\theta)$.
Nevertheless, the $r \phi$ component assumption may work well for
moderate $\theta$: $\pi/4 < \theta < 3\pi/4$. The fact that
the surface condition could not be matched in the range
$\pi/4 < \theta < 3\pi/4$ indicates that the half-opening angle of the disk
($\pi/2 - \theta_0$) is likely to be larger than $\pi/4$.

The profile of $c_{\rm s}$ in Figure~2(a) is quite different from
that in the previous works with a vertical polytropic assumption
\citep[e.g.,][]{Hoshi77,Gu09}. Under the polytropic relation
$p_{\rm tot} = {\mathcal K} \rho^{1+1/N}$ (normally $1.5 \le N \le 3$, and
for radiation pressure-dominated case, $p_{\rm tot}$ can be replaced
by $E/3$), $c_{\rm s}$ will decrease continuously from the equatorial plane
to the surface. The reason for the opposite behavior of $c_{\rm s}$,
as implied in Figure~2(a),
is that $\rho$ drops faster than $E$ from the equatorial plane to the surface.
Figure~3 shows the variation of the quantity $d\ln E / d\ln \rho$
with $\theta$ for $\dot m = 0.5$ (dashed line), $\dot m = 0.6$ (solid line),
and $\dot m = 1$ (dotted line). If the polytropic relation works well,
$d\ln E / d\ln \rho$ should be a constant of $1+1/N$. It is clearly shown
in Figure~3 that, however, $d\ln E / d\ln \rho$ varies significantly with
$\theta$ rather than being a constant. More importantly,
$d\ln E / d\ln \rho < 1$ indicates that $N$ is negative thus unacceptable.
We therefore argue that the polytropic relation should be unsuitable for
describing the vertical structure of radiation pressure-supported disks.
Moreover, since the energy advection is relevant to $c_{\rm s}$
\citep[e.g., $Q_{\rm adv} \simeq \dot M c_{\rm s}^2/2 \pi R^2$ in]
[where $Q_{\rm adv}$ is the advective cooling rate per unit area]{Abram95},
we may expect essentially different results on the strength of advection.

\subsection{Energy advection}

Figure~4 shows the variation of the vertically averaged advection factor
$f_{\rm adv}$ with the mass accretion rate $\dot m$, where $f_{\rm adv}$
is defined as $f_{\rm adv} \equiv Q_{\rm adv} / Q_{\rm vis}$. The quantities
$Q_{\rm adv}$ and $Q_{\rm vis}$ are expressed as follows:
\begin{equation}
Q_{\rm adv} = r \int_{\theta_0}^{\pi - \theta_0} q_{\rm adv}
\sin\theta \ d\theta \ ,
\end{equation}
\begin{equation}
Q_{\rm vis} = r \int_{\theta_0}^{\pi - \theta_0} q_{\rm vis}
\sin\theta \ d\theta \ ,
\end{equation}
where $q_{\rm adv} = - v_r E / 2r$
and $q_{\rm vis} = - 3 \rho v_r v_{\phi}^2 / 2r$
are respectively the advective cooling rate and the viscous heating rate
per unit volume, as implied by the left-hand side of Equation~(11).

The solid line in Figure~4 corresponds to the total accretion rate integrating
from $\theta_0$ to $\pi - \theta_0$, as shown by Equation~(16),
whereas the dashed line corresponds to the specific accretion rate
integrating from $\theta = \pi/4$ to $\theta = 3\pi/4$, i.e.,
\begin{equation}
\dot M_{\pi/4} =  - 2 \pi r^2 \int_{\pi/4}^{3\pi/4}
\rho v_r \sin\theta \ d\theta \ .
\end{equation}
The reason why we calculate for $\dot M_{\pi/4}$ is that the $r \phi$
stress assumption may work well for $\pi/4 < \theta < 3\pi/4$.
As shown by the horizontal range of the solid and dashed lines,
most of the accreted matter exists in this specific
range, e.g., $\dot m_{\pi/4} = 0.52$ corresponding to $\dot m = 0.6$.
The figure also shows that $f_{\rm adv}$ rapidly increases
with increasing $\dot m$ in the range $0.5 \la \dot m \la 1.1$.
More importantly, the value of
$f_{\rm adv}$ ($0.2 \la f_{\rm adv} \la 0.8$) indicates that
the energy advection is significant even for sub-Eddington accretion disks.

Such a result is quite different from the previous one, where advection
was found to be significant only for super-Eddington accretion case.
\citet{Watarai00} introduced an elegant formula to describe
the $\dot M - L$ relationship based on their numerical solutions
under the well-known Paczy\'{n}ski-Wiita potential \citep{PW80}.
Their Equations~(15)-(19) imply that, for the position $r = 10r_{\rm g}$,
advection is negligible for $\dot M \la 67 L_{\rm Edd}/c^2$
($\sim 4 \dot M_{\rm Edd}$). For the whole disk, advection is negligible
for $\dot M \la 20 L_{\rm Edd}/c^2$ ($1.25 \dot M_{\rm Edd}$).
Such a critical $\dot M$ for the whole disk was confirmed by 
some recent global solutions under the general relativity.
Figure~4.11 of \citet{Sadow11a} shows that advection is negligible
for $L \la L_{\rm Edd}$ for any spin parameter $a_*$.
For $a_* = 0$, i.e., the Schwarzschild black hole, the critical
$\dot M$ is just around $\dot M_{\rm Edd}$.
In our opinion, the different results on the advection between the above
two works and ours are related to the different approach to describing
the vertical structure.

A significant difference is that \citet{Watarai00} and \citet{Sadow11b}
chose the cylindrical coordinates whereas we adopt the spherical coordinates.
Of course, the final results should not depend on the coordinates used.
However, as pointed out by \citet{Abram97}, there are some interesting
differences between the equations written in cylindrical and
spherical coordinates.
There is no centrifugal force in the $z$ direction in cylindrical
coordinates, whereas there is no gravitational force in the $\theta$
direction in spherical coordinates. \citet{Abram97} claimed that
it is exactly this property that makes the spherical coordinates
much better adapted for describing the flow near the black hole horizon.
Here, we argue that the spherical coordinates should be more
suitable for describing geometrically thick disks as follows.
In cylindrical coordinates, in the $z$ direction, whether with a polytropic
relation between the pressure and the density \citep{Watarai00},
or with the local energy balance \citep{Sadow11b},
an approximation for the gravitational force, i.e., 
$\partial \psi / \partial z = \Omega_{\rm K}^2 z$, was adopted
for describing the vertical structure. Such an approximation
will probably be invalid for $z/r \ga 1$. In particular for $z \to \infty$,
the approximate force goes to infinity whereas the real force
ought to vanish. Thus, the cylindrical coordinates seem
unsuitable for studying geometrically thick disks. In other words,
a geometrically thick disk solution in cylindrical coordinates
may not be self-consistent.
\citet{Sadow11b} limited their solutions by
$\dot M \leqslant 2 \dot M_{\rm Edd}$ probably due to this reason.
As shown by their Figure~10, the maximal value of $H/r$
for $\dot M = 2 \dot M_{\rm Edd}$ is $\sim 0.4$.
For higher $\dot M$, the value of $H/r$ will be even larger thus
the solution based on the approximate force may be inaccurate.
On the contrary, in spherical coordinates, there is no need to
make approximation for the gravitational force.
The centrifugal force in the $\theta$ direction,
which takes the place of the $z$-direction gravitational force
in cylindrical coordinates,
is derived in this work by solving the vertical differential equations.
Thus, our approach to the vertical structure seems to be more reasonable.

We would agree that the solutions in \citet{Sadow11b} are likely
to be self-consistent since their $H/r$ is significantly less than unity,
in particular for the solutions with $\dot M \la \dot M_{\rm Edd}$.
Then what are the reasons for the quantitative difference in the advection
for $\dot M \la \dot M_{\rm Edd}$ between their solutions and ours?
In our understanding, there exist three possible reasons as follows.
First, as mentioned in Section~3 of \citet{Sadow11b} for their numerical
methods, the vertical structure is derived by a given advection factor
$f_{\rm adv}$ in advance. The value of $f_{\rm adv}$ is probably obtained
by solving the radial structure on the equatorial plane.
Moreover, their $f_{\rm adv}$ is assumed to be uniform in the $z$ direction.
On the contrary, we obtain a varying $f_{\rm adv}$ by solving the vertical
equations. As shown by our Figure~8, $f_{\rm adv}$ increases significantly
with $z$. We can therefore expect that the vertically averaged
$f_{\rm adv}$ at a cylindrical radius will also be significantly larger
than that at $z=0$, which may explain why our $f_{\rm adv}$ is
larger than that in \citet{Sadow11b}. Second, \citet{Sadow11b} assumed
a uniform $v_R$ and $v_{\phi}$ in the $z$ direction, and
used the Keplerian strain to calculate the viscous dissipation.
In our method, however, we include varying $v_r$ and $v_{\phi}$
in the vertical direction, and the viscous dissipation is calculated
based on $v_{\phi}$ instead of $v_{\rm K}$.
Third, \citet{Sadow11b} assumed $v_z = 0$ whereas we have $v_{\theta} = 0$.
As stressed by \citet{Abram97}, since the stationary accretion flows
resemble quasi-spherical flows ($\theta_0 \approx$ constant)
much more than quasi-horizontal	flows ($H \approx$ constant),
$v_{\theta} = 0$ may be a more reasonable approximation than $v_z = 0$.
Moreover, the above three reasons may also be responsible for
the different results in the convective stability,
as will be discussed in Section~5.3.

\subsection{Spin problem for $L \ga 0.3 L_{\rm Edd}$}

As mentioned in Section~1, some works on the black hole spin measurement
showed that the standard thin disk model is likely to be inaccurate for
$L \ga 0.3 L_{\rm Edd}$ \citep[e.g.,][]{McClin06,Straub11}.
One explanation is that the inner disk edge is still located
at the innermost stable circular orbit (ISCO), but its emission
is shaded by the outer disk.
Thus, the inner disk radius obtained from the spectral fitting is not true.
However, \citet{Weng11} showed that the disks in black hole and
neutron star X-ray binaries trace the same evolutionary pattern
for $L \ga 0.3 L_{\rm Edd}$. In addition, for the neutron star system
XTE J1701-462, the boundary emission area maintains nearly constant despite
the varying luminosity of the disk \citep[Figure~17]{Lin09},
which indicates that the neutron star's surface is not shaded.
\citet{Weng11} therefore argued that the inner disk of the black hole system
should not be shaded either due to the similar phenomenon. They suggested
that the inner disk radius moves outward because of the increasing
radiation pressure.

In our opinion, from the energy advection, it is easy to understand
that the standard disk model seems to be inaccurate above $0.3 L_{\rm Edd}$.
As revealed by the lines in Figure~4, $f_{\rm adv}$ is likely to
be non-negligible (probably $\sim 0.1$) for $\dot m \sim 0.3$
at $r = 10 r_{\rm g}$. We would point out that, compared with   
the Paczy\'{n}ski-Wiita potential, the Newtonian potential
in the present work may magnify the viscous heating rate
at small radii such as $10 r_{\rm g}$,
thus the real $f_{\rm adv}$ at $10 r_{\rm g}$ may be smaller than the
values showed in Figure~4. On the other hand, for the
same $\dot m$, since the viscous heating rate at a smaller radius such as
$r = 5r_{\rm g}$ will probably be larger than that at $r = 10r_{\rm g}$,
so does the advection factor.
We can therefore expect that, even for the Paczy\'{n}ski-Wiita potential,
the advection at the position close to the ISCO should be non-negligible
for $\dot m \sim 0.3$.
Consequently, the standard thin disk model, based on the energy balance
between the viscous heating and the radiative cooling with
the advective cooling being ignored, may be inaccurate.

\subsection{Vertical height}

Figure~2 shows that $\rho$ decreases significantly with decreasing $\theta$,
and Figure~4 implies that most of the accreted matter exists in
the range $\pi/4 < \theta < 3\pi/4$. In order to have a more clear view,
we define an averaged dimensionless height as
$\Delta\theta \equiv \Sigma /2 r \rho_0$, where the surface density $\Sigma$
takes the form:
\begin{equation}
\Sigma = r \int_{\theta_0}^{\pi-\theta_0} \rho \sin\theta \ d\theta \ .
\end{equation}
Figure~5 shows the variation of $f_{\rm adv}$ with $\Delta\theta$ (solid line).
Even though the photosphere is close to the polar axis,
the averaged height $\Delta\theta$ is geometrically slim with
$0.3 \la \Delta\theta \la 0.6$.
Furthermore, the figure shows that $f_{\rm adv}$ increases with
increasing $\Delta\theta$ or $\dot m$, which agrees with the classic picture.
For quantitative comparison, we plot the function
$f_{\rm adv} = 1.5 \tan^2(\Delta\theta)$ (dashed line)
in Figure~5 due to the relationship $f_{\rm adv} \ga (H/R)^2$
introduced by \citet{Abram95}, which is equivalent to
$f_{\rm adv} \ga \tan^2(\Delta\theta)$ here. It is seen that
$f_{\rm adv}$ is not well proportional to $\tan^2(\Delta\theta)$.
In the range $0.6 < \dot m < 1.1$ or $0.3 < f_{\rm adv} < 0.8$, however,
we may regard the formula $f_{\rm adv} = 1.5 \tan^2(\Delta\theta)$
as a rough approximation.

\section{Two-dimensional solutions and convective stability}

\subsection{Two-dimensional solutions}

In Section~4, we focus on the solutions at a typical radius
$r = 10 r_{\rm g}$. In this section we will study the disk solutions
for various radii. Since the vertical solutions are based on
the assumptions of partial derivatives in
the radial direction (presented in Section~2.1), it is necessary to
derive vertical solutions for various radii to check whether these
assumptions are self-consistent.
Following the example solution in Figure~2, we study the two-dimensional
solutions for $\dot m = 0.6$ in the range $6r_{\rm g} \leqslant r
\leqslant 12r_{\rm g}$ and $0 < \theta \leqslant \pi/2$.
Figure~6 shows the radial variations of $c_{\rm s}$ and $E$ (solid lines)
for five polar angles, i.e., $\theta = 90^{\circ}$, $75^{\circ}$,
$60^{\circ}$, $45^{\circ}$, and $30^{\circ}$. For comparison, the radial
profile of $v_{\rm K}$, which is proportional to $r^{-1/2}$,
is shown in Figure~6(a),
and an example slope of $\propto r^{-5/2}$ is shown in Figure~6(b).
The figure shows that $c_{\rm s}$ and $E$ behave roughly as $\propto r^{-1/2}$
and $\propto r^{-5/2}$, respectively, which agrees with the original
assumptions of radial derivatives.
As mentioned in Section 2.1, once the
radial derivatives of $c_{\rm s}$ and $E$ are given, the other
ones can be inferred from Equations (1)-(7).
Thus, we can expect that the radial derivatives of $v_r$, $v_{\phi}$,
$\rho$, and $F_r$ in the two-dimensional solutions should also be
in agreement with the assumptions.
Our solutions are therefore likely to be self-consistent.

In our calculation, the location of the photosphere does not vary much
with the radius, i.e., $\theta_0 \la 5^{\circ}$ for various radii.
As discussed in Section~4.1, the real position of the photosphere
is likely to be located in the range $5^{\circ} < \theta_0 < 45^{\circ}$.
We will make some comparison with simulations for the photosphere in Section~6.
As shown in Figure~6, the derivatives of $c_{\rm s}$ and $E$ 
deviate a little for $r \to 12 r_{\rm g}$ and $\theta = 90^{\circ}$.
Such a divergence may be well understood from Figure~1, which shows that
the solution for $\dot m = 0.6$ and $r \to 12 r_{\rm g}$ is quite close to
the lower solid line, which indicates that the gas pressure may not
be negligible. In particular for the equatorial plane, the mass density
has the maximal value there, thus the gas pressure
may be most significant at this position.

\subsection{Solberg-H\o{}iland conditions}

In this section,
we will study the convective stability of the radiation
pressure-supported disks. The well-known Solberg-H\o{}iland conditions
in cylindrical coordinates ($R$, $\phi$, $z$) take the forms
\citep[e.g.,][]{Tassoul00}:
\begin{equation}
\frac{1}{R^3} \frac{\partial l^2}{\partial R}
- \frac{1}{C_P \rho} {\bma \nabla} P \cdot {\bma \nabla} S > 0 \ ,
\end{equation}
\begin{equation}
- \frac{\partial P}{\partial z}
\left(
\frac{\partial l^2}{\partial R}\frac{\partial S}{\partial z}
- \frac{\partial l^2}{\partial z}\frac{\partial S}{\partial R}
\right) > 0 \ ,
\end{equation}
where $l$ is the specific angular momentum per unit mass, $P$ is the total
pressure, $C_P$ is the specific heat at constant pressure,
and $S$ is the entropy expressed as
\begin{equation}
dS \propto d \ln \left( \frac{P}{\rho^{\gamma}} \right) \ ,
\end{equation}
where $\gamma$ is the adiabatic index.

The $R$ and $z$ components of the well-known
Brunt-V\"ais\"al\"a frequency are written as
\[
N_R^2 = - \frac{1}{\gamma \rho} \frac{\partial P}{\partial R} 
\frac{\partial}{\partial R}\ln \left( \frac{P}{\rho^{\gamma}} \right) \ , 
\]
\[
N_z^2 = - \frac{1}{\gamma \rho} \frac{\partial P}{\partial z}
\frac{\partial}{\partial z}\ln \left( \frac{P}{\rho^{\gamma}} \right) \ ,
\]
and the epicyclic frequency takes the form:
\[
\kappa^2 = \frac{1}{R^3} \frac{\partial l^2}{\partial R} \ .
\]
Thus, the first Solberg-H\o{}iland condition, Equation~(21), can be
simplified as
\begin{equation}
N_{\rm eff}^2 \equiv N_R^2 + N_z^2 + \kappa^2 > 0 \ ,
\end{equation}
where $N_{\rm eff}$ is defined as an effective frequency.
For accretion disks, there usually exists $\partial P / \partial z < 0$
(as shown in Figure~8), so the second Solberg-H\o{}iland condition,
Equation~(22), reduces to
\begin{equation}
\Delta_{lS} \equiv
\frac{\partial l^2}{\partial R}\frac{\partial}{\partial z}
\ln \left( \frac{P}{\rho^{\gamma}} \right) 
- \frac{\partial l^2}{\partial z}\frac{\partial}{\partial R}
\ln \left( \frac{P}{\rho^{\gamma}} \right)
> 0 \ .
\end{equation}
In numerical calculation, we adopt $P = E/3$ and $\gamma = 4/3$
according to the radiation pressure-supported assumption.

\subsection{Convective stability}

Based on the two-dimensional solutions for $\dot m = 0.6$ in Section~5.1,
we can obtain the variations of physical quantities in cylindrical
coordinates and therefore investigate the convective stability
by Equations~(24)-(25).
We take the cylindrical radius $R = 10 r_{\rm g}$ as a typical
position to study the convective stability.

Figure~7 shows the $z$-direction variations of $\kappa^2$, $N_R^2$, $N_z^2$,
$N_{\rm eff}^2$, and $\Delta_{lS}$, where the former four
quantities are normalized by $\Omega_{\rm K}^2$, and the last one is
normalized by $v_{\rm K}^2$. The positive values for both $N_{\rm eff}^2$
and $\Delta_{lS}$ indicate that the disk should be convectively stable.
For the equatorial plane, the result of $N_{\rm eff}^2 > 0$ can be inferred
by Equation~(15) of \citet{NY94}, which revealed that the disk
will always be convectively stable for $\gamma = 4/3$ at $z=0$.

For further understanding the convectively stable results for $z>0$, we plot
Figure~8 to show the $z$-direction variations of $\rho$, $E$, $l$,
$E/\rho^{4/3}$, and the advection factor $f_{\rm adv}$.
It is seen that $\rho$ drops faster than $E$ with increasing $z$,
so the quantity $E/\rho^{4/3}$ increases with $z$. From Equation~(23) we
immediately have
\begin{equation}
\frac{\partial S}{\partial z} \propto
\frac{\partial}{\partial z} \ln \left( \frac{E}{\rho^{4/3}} \right) > 0 \ ,
\end{equation}
which is known as the Schwarzschild criterion for 
a constant angular velocity at a cylindrical surface
($\partial \Omega / \partial z = 0$), corresponding to the so-called
barytropic flows where the pressure depends only on the density.
As the dashed line in Figure~8 shows, the angular momentum $l$
(or equivalently the angular velocity $\Omega$)
does not vary significantly with $z$. Thus, the convectively stable
results are easy to understand from $\partial S / \partial z > 0$.

In a similar study (vertical structure based on the local energy balance)
of the general model for optically thick disks, however,
the disk was found to be convectively unstable
\citep[e.g.,][]{Sadow09, Sadow11b}.
As interpreted by the three possible reasons in Section~4.2, the significant
difference in the results between their works and ours is probably
related to the different approach to describing the vertical structure.
In addition, here we would mention two more details which may help to
understand the difference in convective stability.
First, the profiles of $\rho$ and $E$ in Figure~8 reveal that the absolute
value of radial velocity $|v_r|$ increases with $z$ ($|v_r| \propto E/\rho$
inferred from Equations~(7) and (10)). Compared with the uniform $v_R$
in \citet{Sadow11b}, our increasing $|v_r|$ with $z$ may result
in faster drop of $\rho$ in the $z$ direction (steeper slope of $\rho$)
if we simply assume the mass supply to be comparable.
Second, compared with \citet{Sadow11b}, the advection in our solutions
is significantly stronger, which means that for the same $\dot m$ thus
comparable viscous heating rate, the vertical radiation flux $F_z$ will
be less in our results. Thus, $E$ may decrease slower in the $z$ direction
(flatter slope of $E$) due to the less $F_z$ and lower $\rho$
(inferred from Equation~(6)).
As indicated by Equation~(26), the flatter slope of $E$ and the steeper slope
of $\rho$ will both make contribution to $\partial S / \partial z > 0$,
and the disk is therefore likely to be convectively stable.

We would stress that, our solutions are limited by the radiation
pressure-supported case. Thus, the present results cannot directly
show the convective stability of disks for either the gas pressure-supported
case or the case of comparable gas and radiation pressure.
Actually, we have made some additional calculation for thin disks
in cylindrical coordinates to check the convective stability,
following the method of \citet{Sadow11b} but without considering advection.
We found that the disk
is convectively stable for the gas pressure-supported case, whereas the disk
is convectively unstable for the case of radiation pressure being significant.
Thus, we would agree with \citet{Sadow11b} on the convectively unstable disks
for moderate accretion rates such as $0.01 \sim 0.1 \dot M_{\rm Edd}$,
corresponding to significant radiation pressure
and non-negligible gas pressure.
Moreover, for $\dot m \la 0.1$, the disk will be geometrically thin, thus
there is no difference between the assumptions $v_{\theta} = 0$
and $v_z = 0$, and $f_{\rm adv}$ is probably negligible.
As a consequence, the solutions of \citet{Sadow11b} ought to be accurate.

Furthermore, as revealed by the profiles of $E$ and $\rho$ in Figure~8,
$d\ln E / d\ln \rho$ is less than unity in the $z$ direction.
Following the argument in Section~4.1,
the polytropic relation seems unsuitable either in the $z$ direction.
In addition, as shown in Figures~7 and 8, our example solution
for $\dot m = 0.6$ at $R = 10 r_{\rm g}$ is terminated at $z/R = 0.66$.
The reason is that the solutions in spherical
coordinates is limited by $r = 12 r_{\rm g}$ (as shown by the horizontal
dot-dashed line in Figure~1), which corresponds to $z/R = 0.66$
at $R = 10 r_{\rm g}$ in cylindrical coordinates.

\subsection{Analysis of convective stability for $z \ll R$}

For the region close to the equatorial plane, we can make some analysis
of the convective stability by the Taylor expansion method.
Obviously, we have $\partial S / \partial z = 0$ at $z=0$
from symmetric conditions. Thus, for $z \ll R$,
the value of $\partial S / \partial z$ can be estimated by
the second-order derivative at $z = 0$:
\begin{equation}
\frac{\partial S}{\partial z} \approx
z \frac{\partial^2 S}{\partial z^2}
\qquad (z \ll R) \ .
\end{equation}

Based on Equations~(7)-(12), we can eliminate $v_r$, $E$, and $F_{\theta}$
and therefore obtain a set of three equations for the three quantities
$v_{\phi}$, $c_{\rm s}$, and $\rho$. By using the Taylor
expansion method, together with the boundary condition of Equation~(14),
we derive the following three relationships for the second-order
derivatives of $v_{\phi}$, $c_{\rm s}$, and $\rho$
(the ram-pressure term in Equation~(8) is ignored):
\[
\frac{5 c_{\rm s}}{2} \frac{\partial^2 c_{\rm s}}{\partial \tilde\theta ^2}
+ v_{\phi} \frac{\partial^2 v_{\phi}}{\partial \tilde\theta ^2} = 0 \ ,
\]
\[
\frac{1}{\rho} \frac{\partial^2 \rho}{\partial \tilde\theta ^2}
+ \frac{2}{c_{\rm s}} \frac{\partial^2 c_{\rm s}}{\partial \tilde\theta ^2}
= - \frac{v_{\phi}^2}{c_{\rm s}^2} \ ,
\]
\[
\frac{3}{v_{\phi}} \frac{\partial^2 v_{\phi}}{\partial \tilde\theta ^2}
= \frac{1}{2\rho} \frac{\partial^2 \rho}{\partial \tilde\theta ^2}
+ \frac{1}{c_{\rm s}} \frac{\partial^2 c_{\rm s}}{\partial \tilde\theta ^2}
+ \frac{1}{v_{\phi}^2 - c_{\rm s}^2}
\left( v_{\phi} \frac{\partial^2 v_{\phi}}{\partial \tilde\theta ^2}
- c_{\rm s} \frac{\partial^2 c_{\rm s}}{\partial \tilde\theta ^2} \right) \ ,
\]
where $\tilde\theta$ is defined as $\tilde\theta = \pi/2 - \theta$, which
is a small value in the analysis.

In our solutions we have $c_{\rm s}^2 \ll v_{\phi}^2$ at $z=0$,
e.g., $c_{\rm s}^2 / v_{\phi}^2 = 0.16$ for $\dot m = 0.6$ at
$R=10 r_{\rm g}$. For the simple case with $c_{\rm s}^2 / v_{\phi}^2 \ll 1$,
the above three relationships provide
\[
\frac{\partial^2 \ln v_{\phi}}{\partial \tilde\theta ^2}
\approx - \frac{5}{16} \frac{v_{\phi}^2}{c_{\rm s}^2} \ ; \qquad
\frac{\partial^2 \ln c_{\rm s}}{\partial \tilde\theta ^2}
\approx \frac{1}{8} \frac{v_{\phi}^4}{c_{\rm s}^4} \ ; \qquad
\frac{\partial^2 \ln \rho}{\partial \tilde\theta ^2}
\approx - \frac{1}{4} \frac{v_{\phi}^4}{c_{\rm s}^4} \ .
\]
The second-order derivative of entropy at $z =0$ can therefore
be derived by the following coordinate transformation:
\begin{equation}
\frac{\partial^{2} S}{\partial z^{2}} \propto
\frac{\partial^2}{\partial z^2} \ln
\left( \frac{c_{\rm s}^2}{\rho^{1/3}} \right)
= \frac{1}{R^2} \frac{\partial^2}{\partial \tilde\theta ^2} \ln
\left( \frac{c_{\rm s}^2}{\rho^{1/3}} \right)
+ \frac{1}{R} \frac{\partial}{\partial r} \ln
\left( \frac{c_{\rm s}^2}{\rho^{1/3}} \right)
\approx \frac{1}{R^2} \left( \frac{1}{3} \frac{v_{\phi}^4}{c_{\rm s}^4}
- \frac{1}{2} \right) > 0 \ .
\end{equation}
Thus, Equations (27)-(28) indicate $\partial S / \partial z > 0$ for
the region close to the equatorial plane. The disk in this region
is therefore likely to be convectively stable.

\section{Summary and discussion}

In this paper, we have studied the vertical structure, energy
advection, and convective stability of radiation pressure-supported disks
in spherical coordinates.
In the $\theta$ direction, we replaced the pressure-density polytropic
relation by the local energy balance per unit volume between
the viscous heating and the advective cooling plus the radiative cooling,
and obtained the distribution of physical quantities such
as $\rho$, $v_r$, $v_{\phi}$, $c_{\rm s}$, $E$, and $F_{\theta}$.
The photosphere was found close to the polar axis
and therefore the disk seems to be extremely thick.
However, most of the accreted matter exists in a moderate range
around the equatorial plane such as $\pi/4 < \theta < 3 \pi/4$.
We showed that the polytropic relation is unsuitable for describing
the vertical structure of radiation pressure-supported disks.
More importantly, we found that the energy advection is significant
even for slightly sub-Eddington accretion disks,
which is quite different from
the previous result that the advection is of importance only for
super-Eddington accretion disks. We argued that, the non-negligible
advection may help to understand why the standard thin disk model
is likely to be inaccurate for $L \ga 0.3 L_{\rm Edd}$.
In addition, we studied the two-dimensional solutions to check
our basic assumptions of radial derivatives,
which indicates our solutions to be self-consistent.
Furthermore, we investigated the convective stability of the disks
in cylindrical coordinates by
the two-dimensional solutions derived in spherical coordinates.
The disk solutions satisfy the Solberg-H\o{}iland
conditions, which reveals that the disk ought to be convectively stable.

To our knowledge, there are mainly two series of simulations on
optically thick accretion flows. One is three-dimensional, radiation
magnetohydrodynamic
(RMHD) simulations by a shearing box, which focused on thin
disks \citep[e.g.,][]{Hirose06,Krolik07,Hirose09a,Hirose09b}.
The other is two-dimensional, either radiation hydrodynamic (RHD)
or RMHD simulations for global flows, which included thin disks
and super-Eddington accretion flows
\citep[e.g.,][]{Ohsuga05,Ohsuga09,Ohsuga11}.
We would like to compare our numerical results with the latter
owing to the global, radiation pressure-supported, and geometrically
not thin case.
\citet{Ohsuga05} studied RHD simulations in spherical coordinates
with the $\alpha$ stress assumption for only the $r \phi$ component,
for extremely high accretion rates
$\dot M = 300$, $1000$, and $3000 L_{\rm Edd}/c^2$.
Moreover, \citet{Ohsuga11} studied RMHD simulations
in cylindrical coordinates for $\dot M \sim 100 L_{\rm Edd}/c^2$.
In our results, the position of the photosphere is quite
close to the polar axis with $\theta_0 \la 5^{\circ}$.
As discussed in Section~4.1, the real position may exist in the range
$5^{\circ} < \theta_0 < 45^{\circ}$.
In the simulations, \citet{Ohsuga11} showed
that the photosphere is around $z/R = 2.4$ corresponding to
$\theta_0 \sim 23^{\circ}$. \citet{Ohsuga05} did not mention
the position of photosphere, but their Figure~4 for the density distribution
reveals that the photosphere should exist at a certain $\theta_0$
significantly less than $45^{\circ}$. Thus, we would express that,
for accretion rates around and beyond $\dot M_{\rm Edd}$,
the photosphere may exist far from the equatorial plane with
$\theta_0 < 45^{\circ}$.

In addition, we discuss the possible link between our disk model
and ultraluminous X-ray sources.
As shown by the upper solid line in Figure~1, there exists a maximal
accretion rate $\dot m_{\rm max}$ varying with the radius.
We argue that the possible upper limit of the accretion
rate may help to understand why
most ultraluminous X-ray sources (ULXs) are not in thermal dominant state.
As mentioned in the review paper of \citet{Feng11},
there may exist three classes of black holes in ULXs:
normal stellar mass black holes ($\sim 10 M_{\sun}$),
massive stellar black holes ($\la 100 M_{\sun}$),
and intermediate mass black holes ($10^2 - 10^4 M_{\sun}$).
A massive stellar black hole with a moderate super-Eddington accretion
rate seems to account for most sources up to luminosities $\sim$ a few
$10^{40}$ erg s$^{-1}$.
The slim disk model, which is the classic model for super-Eddington
accretion disks, predicts a dominant thermal radiation from the disk.
However, observations have shown that ULXs are not in the thermal
dominant state except only a few sources such as M82 X-1 \citep{Feng10}
and HLX-1 \citep{Davis11}.
In our understanding, the radiation of ULXs may be interpreted
by an optically thick disk with $\dot m \la \dot m_{\rm max}$
plus strong outflows.
The disk will provide a thermal radiation, which is normally not dominant
because of the moderate $\dot m_{\rm max}$.
On the other hand, the outflows may make
contribution to the non-thermal radiation by the bulk motion
Comptonization \citep{TZ98} or through the jet of
the radiation-pressure driven and magnetically collimated outflow
\citep{Ohsuga11}.

\acknowledgments
The author is particularly grateful to Ramesh Narayan for
constructive suggestions and beneficial discussions.
The author also thanks
Ivan Hubeny, Da-Bin Lin, Aleksander S\c{a}dowski, Lijun Gou, Yucong Zhu,
and Shan-Shan Weng for helpful discussions,
and the referee for providing useful comments to improve the paper.
This work was supported by the National Natural Science Foundation of China
under grants 11073015 and 10833002,
the National Basic Research Program (973 Program) of China
under grant 2009CB824800,
and the scholarship from China Scholarship Council under grant 2009835057.

\clearpage

\begin{figure}
\plotone{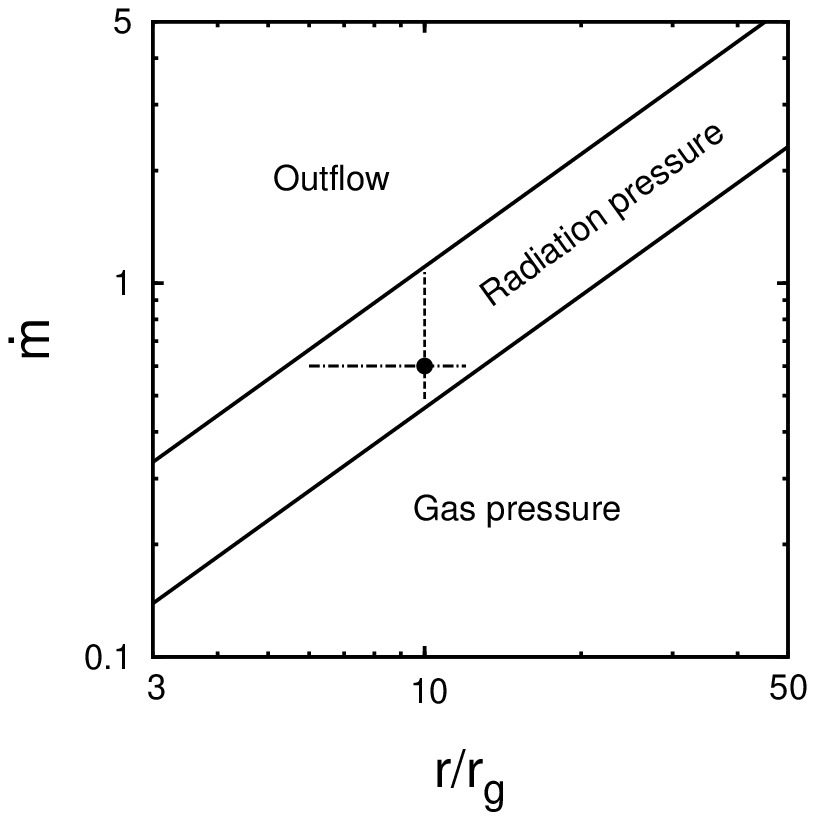}
\caption{
Solutions in the $\dot m$-$r$ diagram. The parameter space is divided into
three regions by two parallel solid lines. The middle region,
denoted by ``radiation pressure", corresponds to the radiation
pressure-supported disk, which is our main interest in this work.
An example solution for $\dot m = 0.6$ at
$r = 10 r_{\rm g}$ (filled circle) is shown in Figure~2.
The solutions for various $\dot m$ at a typical radius $r = 10 r_{\rm g}$
(vertical dashed line) are focused on in Section~4.
The two-dimensional solutions for $\dot m = 0.6$
(horizontal dot-dashed line) are studied in Section~5.
}
\label{fig1}
\end{figure}

\clearpage

\begin{figure}
\plottwo{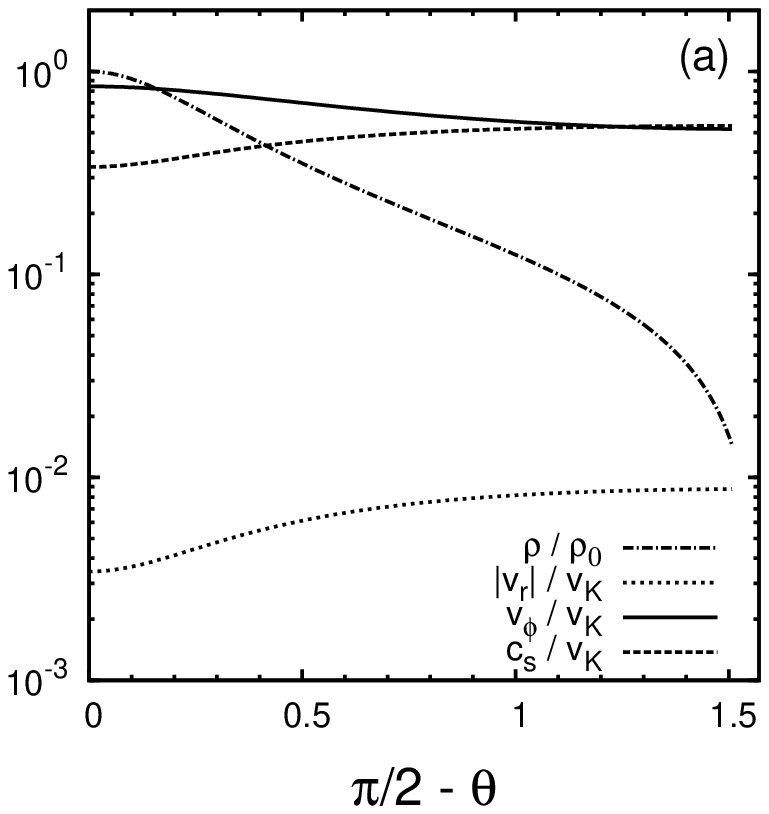}{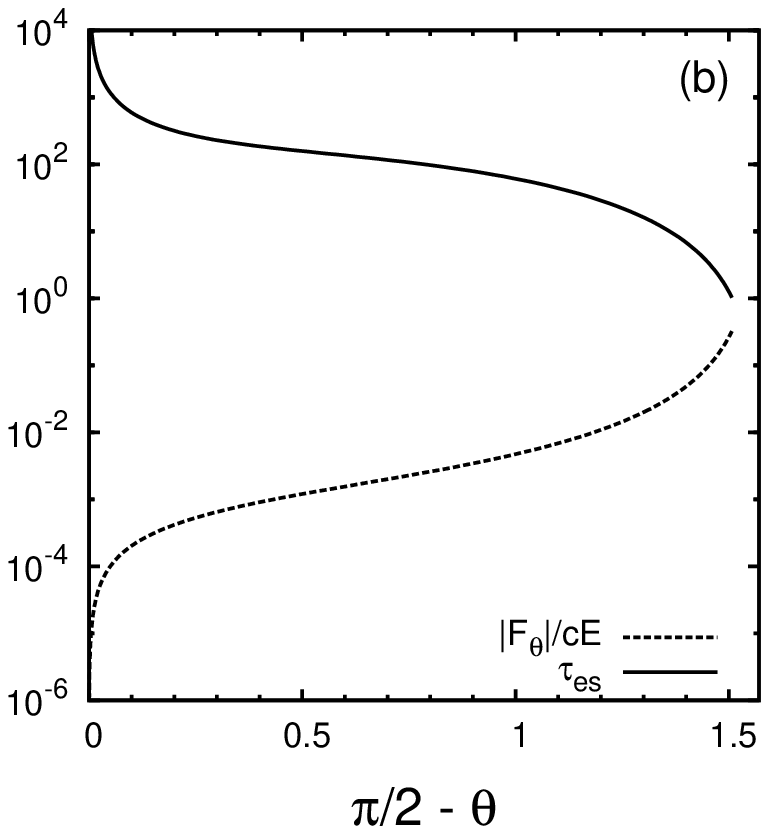}
\caption{
Vertical structure of the disk for $\dot m = 0.6$ at $r = 10 r_{\rm g}$:
(a) variations of $\rho$, $|v_r|$, $v_{\phi}$, and $c_{\rm s}$;
(b) variations of $|F_{\theta}| / cE$ and $\tau_{\rm es}$.
}
\label{fig2}
\end{figure}

\clearpage

\begin{figure}
\plotone{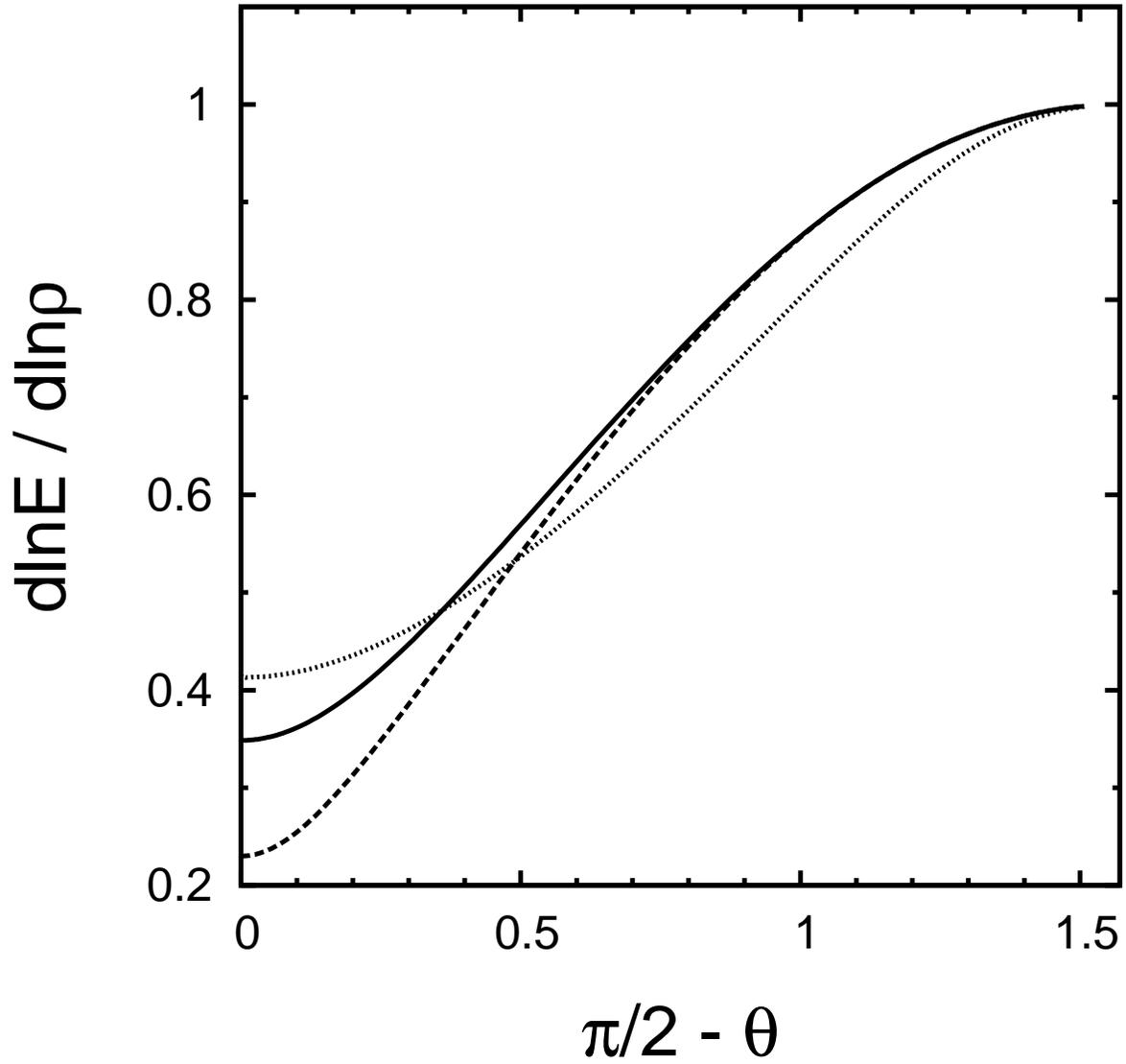}
\caption{
Vertical distribution of $d\ln E / d\ln \rho$ for
$\dot m = 0.5$ (dashed line), $\dot m = 0.6$ (solid line),
and $\dot m = 1$ (dotted line).
}
\label{fig3}
\end{figure}

\clearpage

\begin{figure}
\plotone{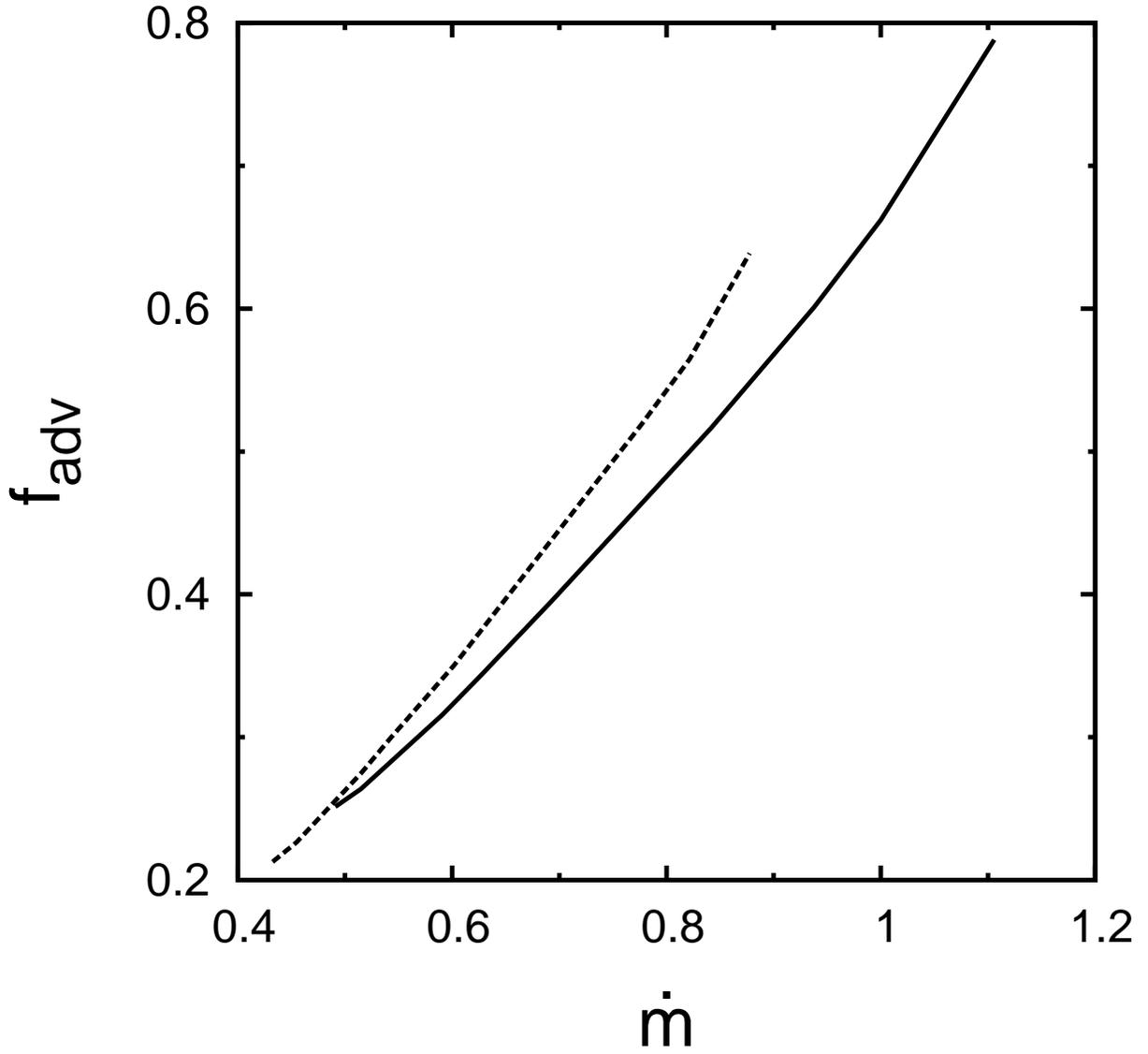}
\caption{
Variation of $f_{\rm adv}$ with $\dot m$ (solid line)
and $\dot m_{\pi/4}$ (dashed line).
}
\label{fig4}
\end{figure}

\clearpage

\begin{figure}
\plotone{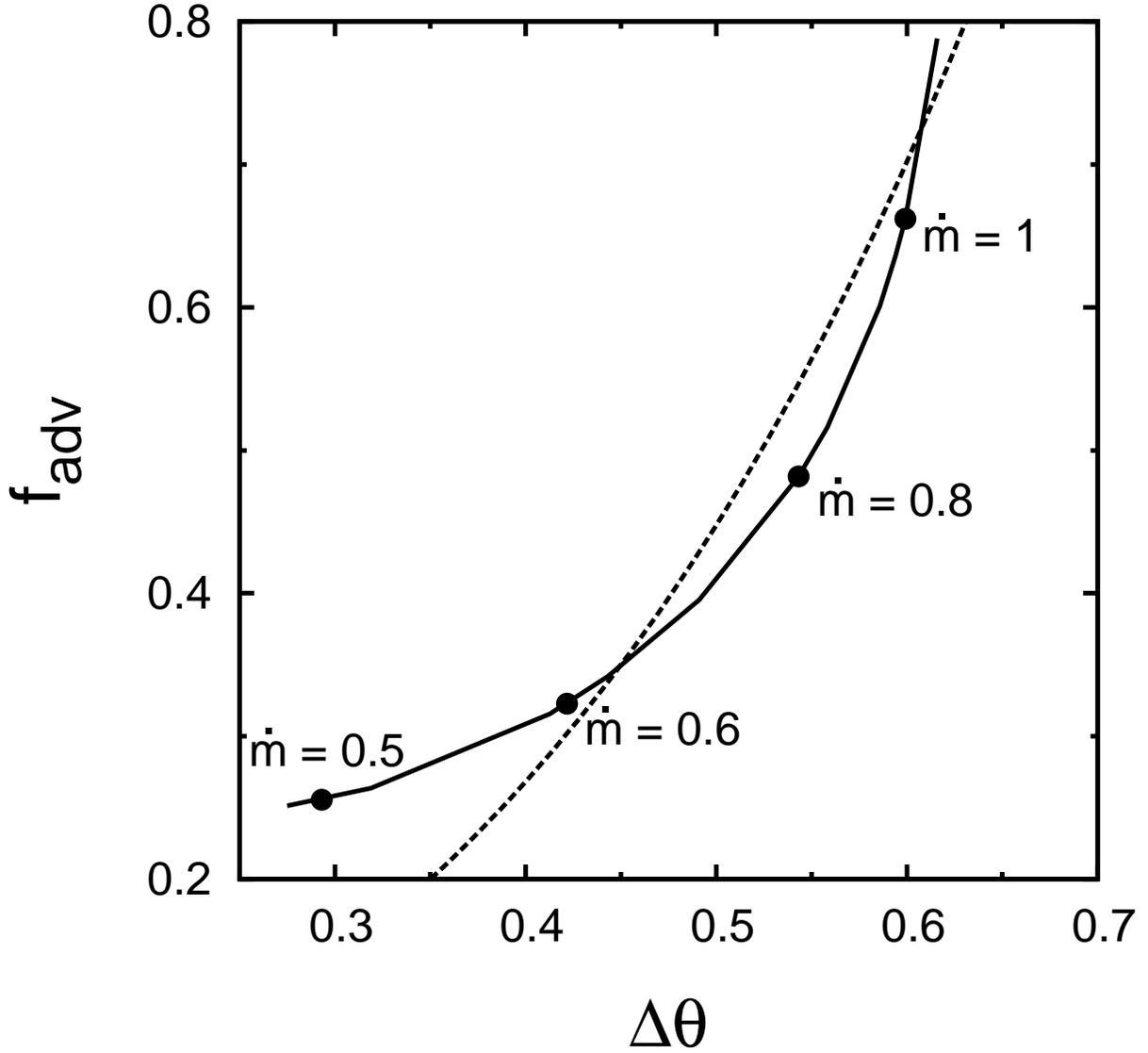}
\caption{
Variation of $f_{\rm adv}$ with the dimensionless height
$\Delta\theta$ (solid line). For comparison,
the function $f_{\rm adv} = 1.5 \tan^2 (\Delta\theta)$ is plotted (dashed line).
The four typical accretion rates, $\dot m = 0.5, \ 0.6, \ 0.8$, and
$1$, are denoted by filled circles.
}
\label{fig5}
\end{figure}

\clearpage

\begin{figure}
\plottwo{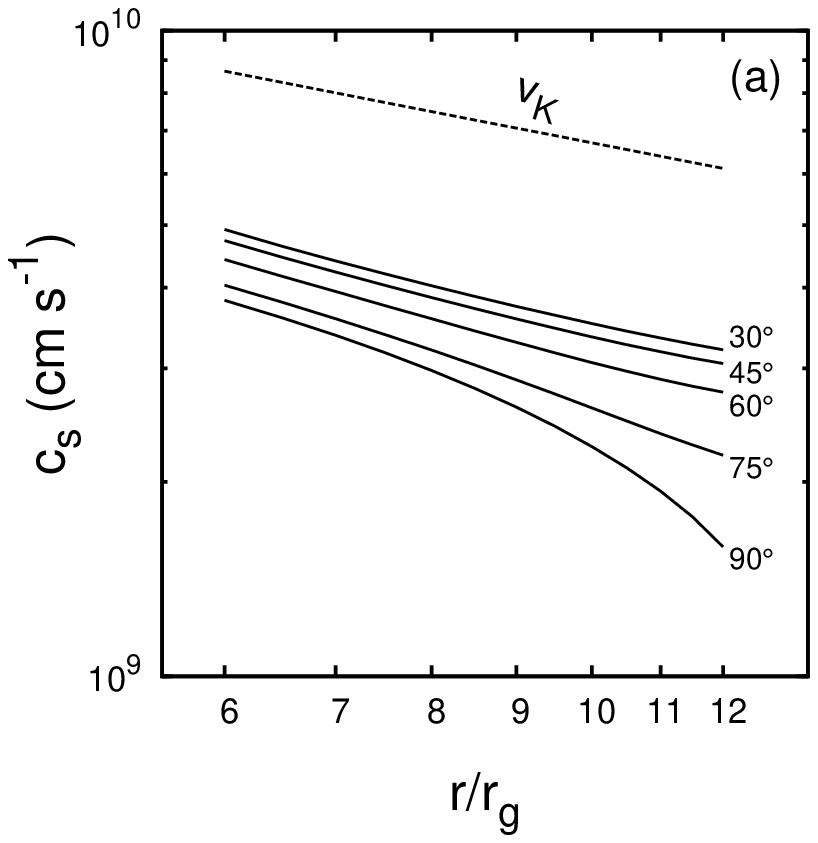}{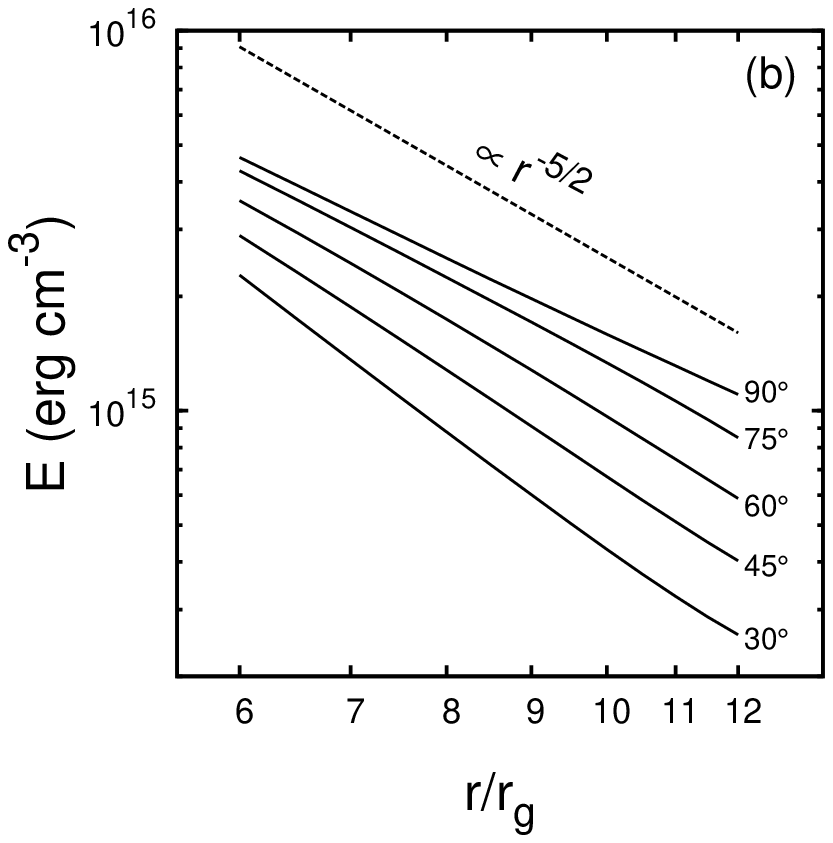}
\caption{
Radial variations of $c_{\rm s}$ and $E$ (solid lines) for the polar angle
$\theta = 90^{\circ}$, $75^{\circ}$, $60^{\circ}$, $45^{\circ}$,
and $30^{\circ}$ for $\dot m = 0.6$.
For comparison, the Keplerian velocity $v_{\rm K}$ ($\propto r^{-1/2}$)
and an example slope of $\propto r^{-5/2}$ are plotted by the dashed lines
in (a) and (b), respectively.
}
\label{fig6}
\end{figure}

\clearpage

\begin{figure}
\plotone{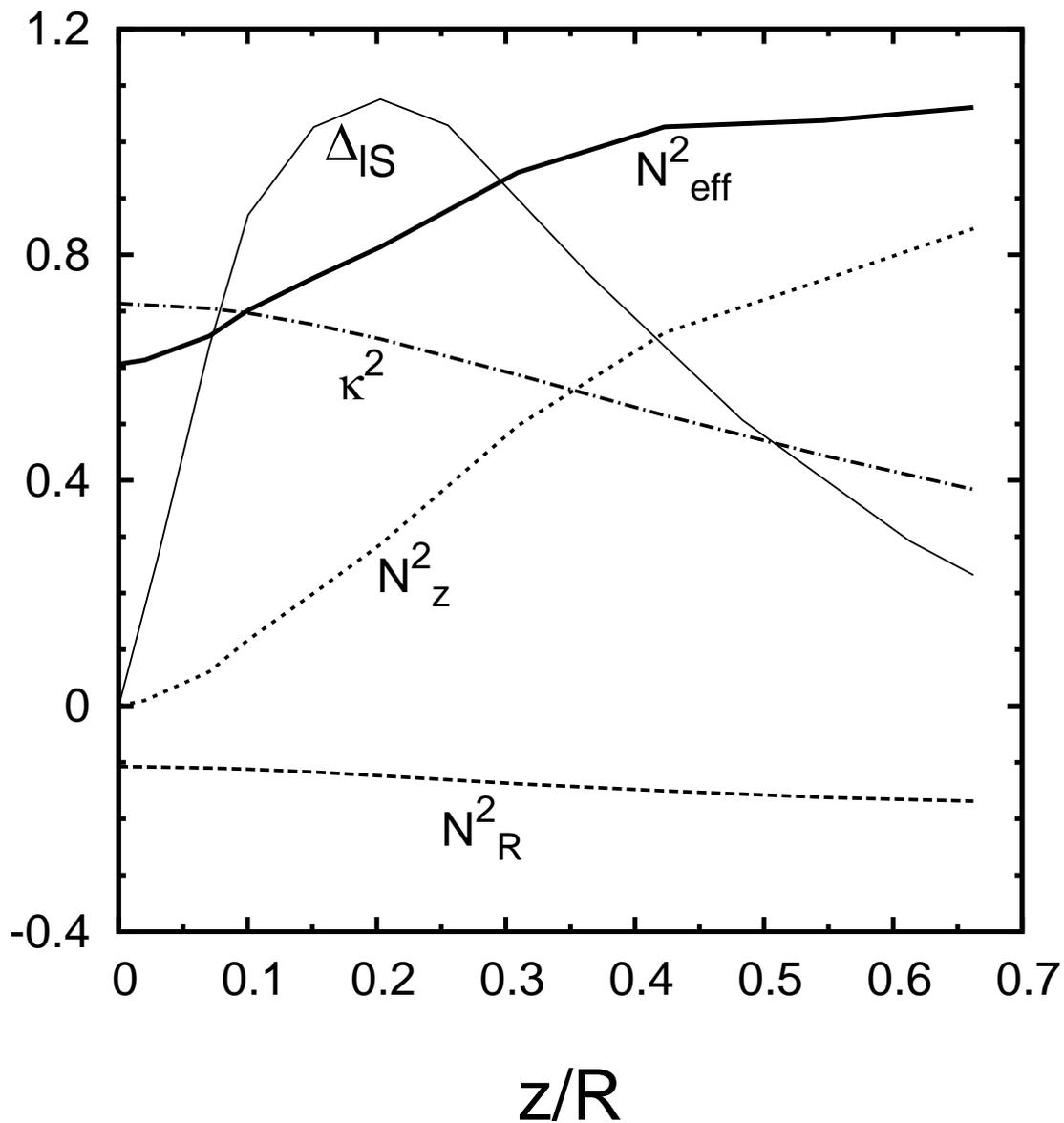}
\caption{
$z$-direction variations of $\kappa^2$ (dot-dashed line),
$N_R^2$ (dashed line), $N_z^2$ (dotted line),
$N_{\rm eff}^2$ (thick solid line), and $\Delta_{lS}$ (thin solid line)
for $\dot m = 0.6$ at a cylindrical radius $R = 10 r_{\rm g}$.
The quantities $\kappa^2$, $N_R^2$, $N_z^2$, and $N_{\rm eff}^2$
are normalized by $\Omega_{\rm K}^2$, and $\Delta_{lS}$ is normalized
by $v_{\rm K}^2$.
}
\label{fig7}
\end{figure}

\clearpage

\begin{figure}
\plotone{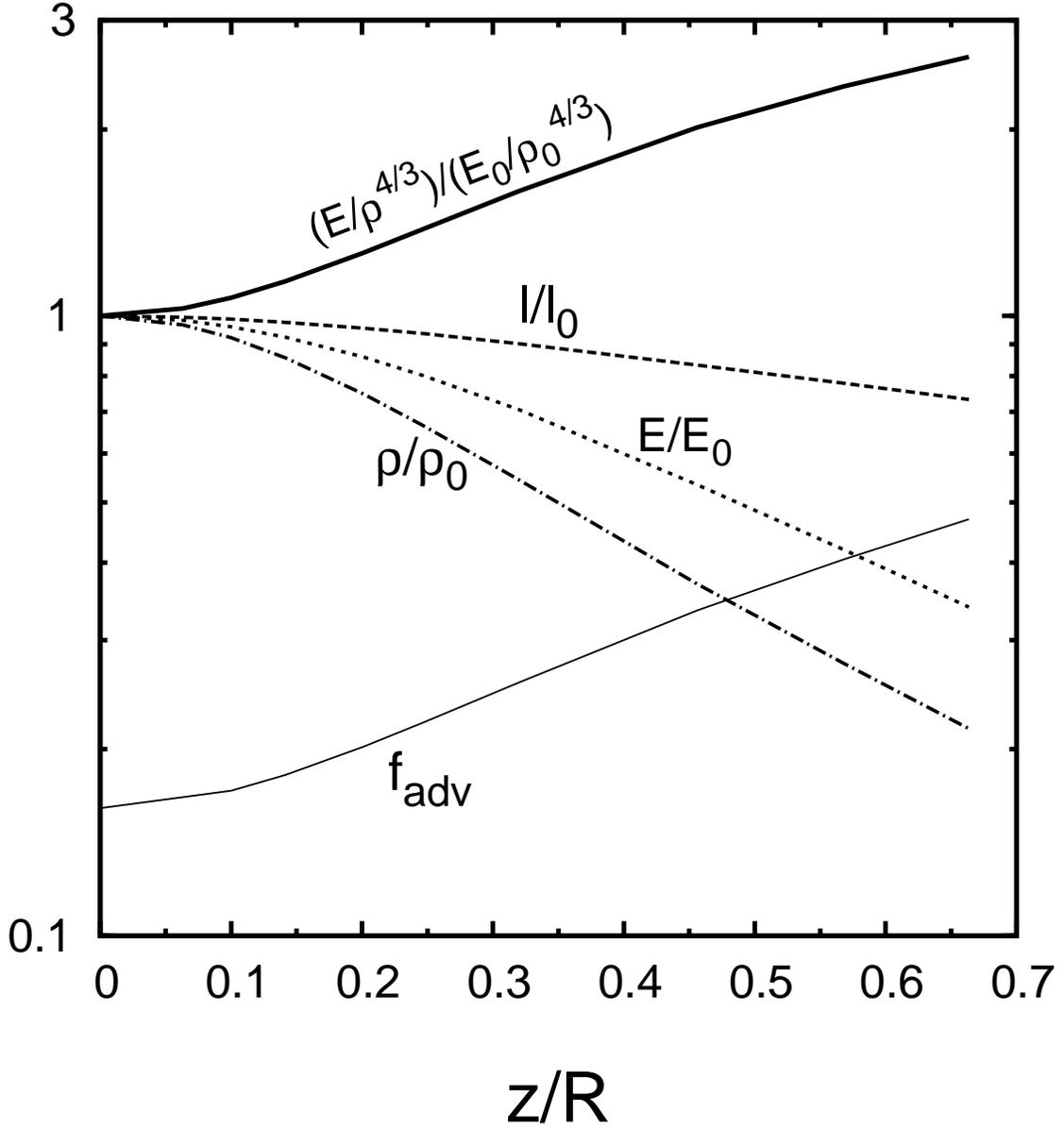}
\caption{
$z$-direction variations of $(E/\rho^{4/3})/(E_0/\rho_0^{4/3})$
(thick solid line),
$l/l_0$ (dashed line), $E/E_0$ (dotted line), $\rho/\rho_0$ (dot-dashed line),
and $f_{\rm adv}$ (thin solid line)
for $\dot m = 0.6$ at $R = 10 r_{\rm g}$,
where the subscript ``0" represents the quantities on the equatorial plane.
}
\label{fig8}
\end{figure}

\end{document}